\documentclass[12pt]{iopart}

\usepackage{amsfonts}
\usepackage{mathrsfs}
\usepackage{color}
\usepackage{hyperref}
\usepackage{cite}
\usepackage{float}
\usepackage{graphicx}
\usepackage{graphics}

\def\C{\mathbb{C}}

\def\H{{\mathcal{H}}}
\date{\empty}
\def\<{\langle}
\def\>{\rangle}
\def\tr{{\rm{tr}}}
\def\i{{\rm{i}}}
\def\e{\rm{e}}

\newcommand{\be}{\begin{equation}}
\newcommand{\ee}{\end{equation}}
\newcommand{\bea}{\begin{eqnarray}}
\newcommand{\eea}{\end{eqnarray}}
\newcommand{\bna}{\begin{eqnarray*}}
\newcommand{\ena}{\end{eqnarray*}}
\renewcommand{\theequation}{\thesection.\arabic{equation}}

\begin{document}
\title[Quantum multipartite maskers vs  quantum error-correcting codes]{Quantum multipartite maskers vs  quantum error-correcting codes}
\author{Kanyuan Han, Zhihua Guo, Huaixin Cao, Yuxing Du, and Chuan Yang}

\address{School of Mathematics and Information Science,  Shaanxi Normal University, Xi'an 710119, China}
\ead{guozhihua@snnu.edu.cn, caohx@snnu.edu.cn}
\vspace{10pt}

\begin{abstract}
Since masking of quantum information was introduced by Modi et al. in [PRL 120, 230501 (2018)], many discussions on this topic have been published.
In this paper, we  explore  the relationship between quantum multipartite maskers (QMMs) and   quantum error-correcting codes (QECCs). We say that a subset $Q$ of pure states of a system  can be masked by an operator $S$ into a multipartite system  if all of the image states $S|\psi\>$ of states $|\psi\>$ in $Q$ have the same marginal states on each subsystem. We call such an $S$ a QMM of $Q$. By establishing a necessary and sufficient for a set  $Q$ be masked by an operator, we explore a relationship between QMMs and QECCs, which reads  that a linear operator is a QMM of all pure states of a system  if and only if its range is a QECC of any one-erasure channel. As an application, we prove  that there is no universal maskers from $\C^2$ into $\C^2\otimes\C^2\otimes\C^2$ and then the states of $\C^3$ can not be masked into $\C^2\otimes\C^2\otimes\C^2$. This gives a consummation to a main result and leads to a negative answer to an open question in [PRA 98, 062306 (2018)]. Another application is that
arbitrary quantum states  of $\C^d$ can be completely hidden
 in correlations between any two subsystems of the tripartite system $\C^{d+1}\otimes\C^{d+1}\otimes\C^{d+1}$, while arbitrary quantum states cannot be completely hidden in the correlations between subsystems of a bipartite system [PRL 98, 080502 (2007)].

\end{abstract}




\section{Introduction}
\setcounter{section}{1}

The importance of information and communication security is increasing rapidly due to the indispensability of internet in now days. Quantum information science (QIS) is an emerging field with the potential to cause revolutionary advances in fields of science and engineering involving computation, communication, precision measurement, and fundamental quantum science. With the repaid development of QIS, many  protocols for quantum communication have been proposed,  including  quantum key distribution \cite{Sas,Schmid}, quantum secret sharing \cite{Schmid,[24]}, and quantum secure direct communication    \cite{Wei,Xie,Zhou1,Zhou2}, which have been widely explored recently. Encoding of quantum information and  error-correcting of quantum channels are fundamental and necessary in quantum communication.
Quantum information is the information that is held in the state of a quantum system and can be manipulated using engineering techniques known as quantum information processing. It was proved that quantum correlation, including Bell nonlocality and  steerability \cite{Bell,Fei,CaoGuo1},
    quantum entanglement \cite{Ent,Yang}, and quantum discord \cite{Luo,GuoZH,Wu,ZhiHua}, is an important resource of quantum information processing.
In quantum mechanics, there are many ``no-go theorems" meaning that to do something according to quantum theory is impossible,
say the no-cloning theorem \cite{Clon[1],Clon[2],Clon[3],Clon[4],Clon[5]}, the no-broadcasting
theorem \cite{broad[5],broad[6]}, the no-deleting theorem \cite{delt1,delt2}, the no-hiding theorem\cite{Brau-Pati,hiding[7]}, which claims that arbitrary quantum states cannot completely
hide in correlations between a pair of subsystems, as well as
no-signalling theorem \cite{sign}. Modi et al. \cite{Mask} discussed the problem of masking quantum information contained in some pure states with a linear operator and obtained the so called no-masking theorem, which says that it is impossible to mask an arbitrary state. It was also proved in  \cite{Mask} that there are sets of nonorthogonal states whose information can be masked. Just as no-go theories being of great significance in information
processing \cite{[13],[14],[18],18}, masking of quantum information has potential applications\cite{[22],[24]}. Li and Wang \cite{Li-Wang} discussed the problem of masking quantum information in multipartite scenario and proved that quantum states can be masked when more
participants are allowed in the masking process. Li et al. \cite{LiBo1}
considered the problem of what kinds of quantum states can be either deterministically or probabilistically masked and proved that mutually orthogonal quantum states can always be served for deterministic masking of quantum information. They also  constructed a probabilistic masking machine for linearly independent states. Liang et al.  \cite{LiBo2} studied the problem of information masking through nonzero linear operators and proved that a nonzero linear operator cannot mask any nonzero measure set of qubit states. They also shown that the maximal maskable set of states on the Bloch
sphere with respect to any masker is the ones on a spherical circle. Furthermore, they given a proof of the conjecture on maskable qubit states proposed by Modi et al. in \cite{Mask}. Moreover, Li and Modi \cite{LiMoa} discussed the problems of probabilistic and approximate masking of quantum
information and the performance of a masking protocol when we are allowed (probabilistic) approximate protocol. They also proved that an $\varepsilon$-approximate universal masker for all states does not exist if the error bound $\varepsilon$ is less than a bound. For more discussions of masking, please refer to references \cite{Ding,LieK,LieJ,Ghosh,CaoGuo}.
Recently, Li and Wang \cite{Li-Wang} discussed masking quantum information in multipartite scenario,
 presented some schemes different from error correction codes, which show that quantum states can be masked when more
participants are allowed in the masking process. They also proved that any all pure  states of the system $\C^d$ can be masked into the tripartite system  $\C^d\otimes\C^d\otimes\C^d$ except for $d=2,6$, using a pair of mutually orthogonal Latin squares of
dimension $d$.

In this paper, we  explore the relationship between quantum multipartite maskers (QMMs) and  quantum error-correcting codes (QECCs). In Sect. 2, we give the definition of a quantum multipartite masker and derive some basic conclusions. In Sect. 3,
we prove a necessary and sufficient condition for an operator to be a QMM and then establish a relationship between QMMs and QECCs. As an application, we prove  that there is no universal masker from $\C^2$ into $\C^2\otimes\C^2\otimes\C^2$, leading to a consummation to the Li and Wang's result mentioned above. We also prove that it is impossible to mask all pure states of any system $K$ into any bipartite system $H_1\otimes H_2$, which generalizes the known no-masking theorem.
\section{Definition and questions}

We use notations $PS(H)$ and $D(H)$ to denote the sets of all pure states  and all mixed states of a quantum system with the state space $H$, respectively. We use $[n]$ to denote the set $\{1,2,\ldots,n\}$ and set $\delta_{x,y}=0(x\ne y)$, $\delta_{x,x}=1.$ Recall that a linear operator $T$ from a Hilbert space $H$ into a Hilbert space $K$ is said to be isometric (or, an isometry) if it is norm-preserving: $\|Tx\|=\|x\|$ for all $x$ in $H$, equivalently, $T^\dag T=I_H$; it is said to be a unitary if it is a surjective isometry,  equivalently, $T^\dag T=I_H$ and $TT^\dag=I_K$.

It was proved in \cite[Theorem 3]{Mask} that an arbitrary quantum state cannot be masked into a bipartite system and pointed out that it is possible to mask an arbitrary quantum state with more than two parties allowed. According to this idea, a generalization of masking defined in \cite{Mask} was proposed in \cite[Definition 1]{Li-Wang}, in which the original system $H_{1}$ is one of the masking participants $H_{1},H_{2},\ldots,H_{n}$.
If the information contained in some states of a system $K$ is first transformed as the information contained in  states of a system $H_{1}$ and then is masked into $H_{1},H_{2},\ldots,H_{n}$, then it is necessary to introduce the following concept.

In what follows, we use the follows notations:

$$\mathcal{H}^{(n)}=\otimes_{j=1}^nH_j=H_{1}\otimes H_{2}\otimes \cdots\otimes H_{n}.$$

{\bf Definition 2.1.} Let $Q$ be a subset of $PS(K)$ and  $S:K\rightarrow \mathcal{H}^{(n)}$ be a linear operator. If there are mixed states $\rho_j\in D(H_j)(j\in[n])$ such that $\forall j\in[n]$, it holds that
\be\label{DD1}
\tr_{\hat{j}}[S|\psi\>\<\psi|S^\dag]=\rho_j, \ \forall |\psi\>\in Q,\ee
where ${\hat{j}}=[n]\setminus\{j\}$, then we say that the information contained in $Q$ can be masked by $S$ into $\mathcal{H}^{(n)}$.
We also say that the operator $S$ is a {\it quantum
multipartite masker (QMM)} (shortly, a masker) for $Q$.
Especially, when $S$ can mask $PS(K)$ into $\mathcal{H}^{(n)}$,  we call it a {\it  universal masker} of $K$ into $\mathcal{H}^{(n)}$, or the system $K$ can be masked into $\mathcal{H}^{(n)}$. Clearly, a universal masker is an isometry since it maps every pure state of $K$ as a pure state of $\mathcal{H}^{(n)}$.

Here are some remarks about Definition 2.1. By definition, an operator from $K$ into $\mathcal{H}^{(n)}$ is a QMM of $Q$ if and only if all of the image states $S|\psi\>$ of states $|\psi\>$ in $Q$ have the same marginal states on each subsystem.
To ensure that operators $\tr_{\hat{j}}[S|\psi\>\<\psi|S^\dag]$ are mixed states of $\mathcal{H}^{(n)}$, it suffices to assume that the operator $S$ is norm-preserving, i.e., it is an isometry: $S^\dag S=I_K$. Clearly, a universal masker must be isometric.
To model physically an isometric masker $S:K\rightarrow \mathcal{H}^{(n)}$ for a set $Q\subset PS(K)$ with a unitary operator $U_S$ on the Hilbert space $\mathcal{H}^{(n)}$, we assume that $\dim(K)\le\dim(H_j)$ for some $j\in[n]$, say $\dim(K)\le\dim(H_1)$. Then we can define an isometry $J:K\rightarrow H_1$ and choose an ancillary state $|b\>$ in $\otimes_{j=2}^nH_j$, and then define an operator $\tilde{S}:J(K)\otimes |b\>\rightarrow {\rm{ran}}(S)$ by
$\tilde{S}(J|\psi\>\otimes|b\>)=S|\psi\>$ for all $|\psi\>$ in $K$.
Since $J$ and $S$ are isometries, $\tilde{S}$ is a unitary operator (i.e. a surjective isometry) and $\dim(J(K)\otimes|b\>)^\perp=\dim({\rm{ran}}(S))^\perp=\dim(\ker(S^\dag))$. Thus, we can choose a unitary operator $V:(J(K)\otimes|b\>)^\perp\rightarrow \ker(S^\dag)$ and define an operator $U_{S,V}:\mathcal{H}^{(n)}\rightarrow \mathcal{H}^{(n)}$ by
$$U_{S,V}=\left(
        \begin{array}{cc}
          \tilde{S} & 0 \\
          0 & V \\
        \end{array}
      \right): (J(K)\otimes|b\>)\oplus (J(K)\otimes|b\>)^\perp)\rightarrow {\rm{ran}}(S)\oplus \ker(S^\dag).$$
Clearly, $U_{S,V}$ is a unitary operator on $\mathcal{H}^{(n)}$ satisfying
$$U_{S,V}(J|\psi\>\otimes|b\>)=\tilde{S}(J|\psi\>\otimes|b\>)=S|\psi\>,\ \ \forall |\psi\>\in Q.$$
This shows that the masker $S:K\rightarrow \mathcal{H}^{(n)}$ for a set $Q\subset PS(K)$ can be modeled by a unitary operator $U_{S,V}:\mathcal{H}^{(n)}\rightarrow \mathcal{H}^{(n)}$ in such a way that
$$S|\psi\>=U_{S,V}(J|\psi\>|b\>),\ \ \forall |\psi\>\in Q.$$

In the case that $K=H_1$ and $n=2d$, Li and Wang proved in \cite[Theorem 1]{Li-Wang} that for any positive integer $d\ge2$, $PS(\C^d)$ can be masked into $(\C^d)^{\otimes2d}$ with the same marginal state  $\frac{1}{d}I_d$.
In the case that $K=H_1$ and $n=3$, the following masking theorem was also established in \cite{Li-Wang}.

{\bf Theorem 2.1}\cite[Corollary 2]{Li-Wang} {\it For all positive integer $d$ larger than $2$ and not equal to $6$, there exists an isometric masker $S_d$ from $\C^d$ into $\C^d\otimes\C^d\otimes\C^d$  with the same marginal state $\frac{1}{d}I_d.$}

The construction of the masker $S_d$ is based on the existence of a pair of orthogonal Latin squares of dimension $d$, and then is very technical and beautiful. The construction of  $S_d$ is as follows.
\be\label{Sd}
S_d|j\>=\frac{1}{\sqrt{d}}\sum_{k=1}^d|k\>|v_{jk}\>|w_{jk}\>(j=1,2,\ldots,d),\ee
where $\{|1\>,|2\>,\ldots,|d\>\}$ is an orthonormal basis (ONB) for $\C^d$, $V=[v_{jk}]$ and $W=[w_{jk}]$ are a pair of orthogonal Latin squares of order $d$. For example, when $d=3$,
$$
V=\left(
    \begin{array}{ccc}
      1 & 2 & 3 \\
      2 & 3 & 1 \\
      3 & 1 & 2 \\
    \end{array}
  \right), \ W=\left(
    \begin{array}{ccc}
      1 & 2 & 3 \\
      3 & 1 & 2 \\
      2 & 3 & 1 \\
    \end{array}
  \right)$$
are pair of orthogonal Latin squares of order $3$, and then we have
$$S_3|1\>=\frac{1}{\sqrt{3}}[|111\>+|222\>+|333\>],$$
$$S_3|2\>=\frac{1}{\sqrt{3}}[|123\>+|231\>+|312\>],$$
$$S_3|3\>=\frac{1}{\sqrt{3}}[|132\>+|213\>+|321\>].$$
Clearly, $S_3$ here is the same as the mapping given by Eq. (1) in \cite{[24]} where the basis $\{|1\>,|2\>,|3\>\}$ was denoted by $\{|0\>,|1\>,|2\>\}$.

With such a masker $S_d$ in Theorem 2.1, we can discuss the masking of quantum information contained in mixed states, i.e.,  the masking of mixed states. For every mixed state $\sigma$ of $\C^d$, we have
$$\sigma=\sum_{k=1}^dc_k|\psi_k\>\<\psi_k|,$$
which is the spectrum decomposition of $\sigma$. Since
$$\tr_{\hat{j}}(S_d\sigma S_d^\dag)=\sum_{k=1}^dc_k\tr_{\hat{j}}(S_d|\psi_k\>\<\psi_k|S_d^\dag)=
\sum_{k=1}^dc_k\rho_j=\rho_j$$
for all $j=1,2,3$, we can say that all of the mixed states of  the system $\C^d$ can be masked into the tripartite system $\C^d\otimes\C^d\otimes\C^d$ by the masker $S_d$ with marginal states $\rho_1,\rho_2$ and $\rho_3$.

Here are extensions of Theorem 2.1 in two directions.

{\bf Corollary 2.1.(Extension of dimensions)} {\it There is a  universal  masker ${S}: K\rightarrow H_1\otimes H_2\otimes H_3$ provided that  $\dim(H_k)\ge\dim(K)\ge3(k=1,2,3)$ and $\dim(K)\ne6$.}

{\bf Proof.} Let $d=\dim(K)$ and let $S_d:\C^d\rightarrow\C^d\otimes\C^d\otimes\C^d$ be a masker with marginal states $\rho_1,\rho_2$ and $\rho_3$. Choose a unitary operator $U:K\rightarrow\C^d$ and an isometry $V_k:\C^d\rightarrow H_k$ for each $k=1,2,3$, and then we get an isometry
$$S=(V_1\otimes V_2\otimes V_3)S_dU:K\rightarrow H_1\otimes H_2\otimes H_3.$$
It is easy to check that $S$ is a masker of pure states of $K$ into $H_1\otimes H_2\otimes H_3$ with the marginal states
$V_1\rho_1V_1^\dag,V_2\rho_2V_2^\dag$ and $V_3\rho_3V_3^\dag$.
The proof is completed.

{\bf Corollary 2.2.(Extension of participants)} {\it When
$n\ge3, \dim(H_k)\ge\dim(K)\ge3(k=1,2,\ldots,n)$, and $\dim(K)\ne6$, there is a universal masker of $K$ into $\mathcal{H}^{(n)}$.}

{\bf Proof.} When $n=3$, the conclusion follows from Corollary 2.1. Next, we assume that $n>3$. Corollary 2.1 implies that there is
an isometric masker $S_0$ of pure states of $K$ into $H_1\otimes H_2\otimes H_3$ with the marginal states
$\eta_1,\eta_2$ and $\eta_3$. Taking pure states $|e_i\>\in\H_i(i=4,5,\ldots,n)$, we obtain an isometry
$A$ from $H_1\otimes H_2\otimes H_3$ into $\mathcal{H}^{(n)}$
satisfying $A|\psi\>=|\psi\>\otimes|e_{4}\>\otimes\cdots\otimes|e_n\>$ for all $|\psi\>$ in $H_1\otimes H_2\otimes H_3$. It is easy to prove that the mapping $AS_0:K\rightarrow \mathcal{H}^{(n)}$
becomes an isometric masker with the marginal states
$\eta_1,\eta_2,\eta_3,|e_4\>\<e_4|,\ldots,|e_n\>\<e_n|.$
The proof is completed.

Furthermore, let $S_k$ be the masker of $PS(\C^k)$ into $\C^k\otimes\C^k\otimes\C^k$ where $k\ne2,6$, and let
$J_{k,k+1}$ be the canonical imbedding of $\C^k$ into $\C^{k+1}$, i.e., $J_{k,k+1}|x\>=|x\>\oplus0$. Then we can see from Theorem 2.1 and Corollary 2.1 that
$PS(\C^k)$ can be masked into $\C^{k+1}\otimes\C^{k+1}\otimes\C^{k+1}$ by using a masker
$\tilde{S}_k$
where
$\tilde{S}_k=S_{k+1}J_{k,k+1}(k\ne5)$, with the same marginal state $\frac{1}{k+1}I_{k+1}(k\ne5)$, and $\tilde{S}_5=(J_{5,6}\otimes J_{5,6}\otimes J_{5,6})S_5$ with the same marginal state $\frac{1}{5}J_{5,6}^\dag I_5J_{5,6}$, respectively.

Thus, to ask the following question is natural:

{\bf Question 2.1.} {\it When $d=2$ or $6$, can $PS(\C^d)$ (or a subset $Q$ of $PS(\C^d)$) be masked into $\C^d\otimes\C^d\otimes\C^d$?}

Although, it was proved in  \cite[Theorem 3]{Mask} a universal masker $H_A\rightarrow H_A\otimes H_B$ does not exist, the following question is needed to be discussed.

{\bf Question 2.2.} {\it For given systems $K, H_1,H_2$, does there  a universal masker $S: K\rightarrow H_1\otimes H_2$ exist?}

At the end of paper \cite{Li-Wang}, the authors proposed the following question:

{\bf Question 2.3.} {\it Can all quantum states of level $d$ be hidden into tripartite quantum system $\C^n\otimes\C^n\otimes\C^n$ with $n<d$ or not?}

In the next section, we will discuss the answers to these questions and give our main results.

\section{Main results}
\setcounter{equation}{0}

Technically and mathematically, we use $T_{j}$ to denote the unitary operator from $\mathcal{H}^{(n)}$ onto $H_j\otimes(\otimes_{k\in \hat{j}}H_k)$ that moves  $j$-th tensor factor to the first one:
$$T_{j}:|h_1h_2\cdots h_{j-1}h_jh_{j+1}\cdots h_n\>\mapsto|h_jh_1h_2\cdots h_{j-1}h_{j+1}\cdots h_n\>,$$
for all $|h_k\>\in H_k(k=1,2,\ldots,n).$
Thus, for all operators $X_i$ on $H_i$, it is easy to check that
\be\label{relation1}
{T_{j}}(X_1\cdots X_{j-1}X_jX_{j+1}\cdots X_n)T_{j}^\dag=
X_jX_1\cdots X_{j-1}X_{j+1}\cdots X_n,
\ee
where $X_1X_2$ stands for $X_1\otimes X_2$, and so on, for short.

{\bf Theorem 3.1.} {\it Let $Q$ be a subset of $PS(K)$ and  $S:K\rightarrow \mathcal{H}^{(n)}$ be a linear operator. Then $Q$ can  be masked by  $S$ if and only if for each $j\in[n]$, there exists a probability distribution (PD) $\mathcal{P}_j=\{c_{ij}\}_{i=1}^{r}$ with $c_{ij}>0$, and an orthonormal set $\mathcal{E}_j=\{|e_{ij}\>\}_{i=1}^{r}\subset PS(H_j)$ such that
\be\label{C2.11}
T_{j}S|\psi\>=\sum_{i=1}^{r}\sqrt{c_{ij}}|e_{ij}\>|f^\psi_{ij}\>\in H_j\otimes\left(\otimes_{k\in \hat{j}}H_k\right), \ \forall |\psi\>\in Q,\ee
where $\mathcal{F}^\psi_j=\{|f_{ij}^{\psi}\>\}_{i=1}^{r}$ is an orthonormal set in  $\otimes_{k\in \hat{j}}H_k$ for every $|\psi\>\in Q$.}

 {\bf Proof.} {\it Necessity.} Let $Q$  be masked by  $S$ and $j\in[n]$. Taking a fixed state $|\psi_0\>\in Q$, we see from  Definition 2.1 that
 \be\label{C2.11-1}
 \tr_{\hat{j}}[S|\psi\>\<\psi|S^\dag]=\tr_{\hat{j}}[S|\psi_0\>\<\psi_0|S^\dag],\ \ \forall |\psi\>\in Q.\ee
Thus, $\forall |\psi\>\in Q$, we have
$$
\tr_{\hat{1}}[T_{j}S|\psi\>\<\psi|S^\dag T_{j}^\dag]=
\tr_{\hat{j}}[S|\psi\>\<\psi|S^\dag]=\tr_{\hat{1}}[T_{j}S|\psi_0\>\<\psi_0|S^\dag T_{j}^\dag].$$
Put $$\rho_j=\tr_{\hat{1}}[T_{j}S|\psi_0\>\<\psi_0|S^\dag T_{j}^\dag]\in D(H_j),$$ then for each $|\psi\>\in Q$, $T_{j}S|\psi\>$ and $T_{j}S|\psi_0\>$ are two purifications of a mixed state $\rho_j$ of system $H_j$. Thus, they have the following Schmidt decompositions:
\be\label{C2.11-3}
T_{j}S|\psi\>=\sum_{i=1}^{r}\sqrt{c_{ij}}|e_{ij}\>|f^\psi_{ij}\>,
T_{j}S|\psi_0\>=\sum_{i=1}^{r}\sqrt{c_{ij}}|e_{ij}\>|g_{ij}\>,
\ee
where $\<f^\psi_{sj}|f^\psi_{tj}\>=\delta_{s,t}$ for all $s,t=1,2,\ldots,r$, $\{|e_{ij}\>\}_{i=1}^{r}$ and $\{|g_{ij}\>\}_{i=1}^r$ are orthonormal sets in $H_j$ and $\otimes_{k\in \hat{j}}H_k$, respectively, which are independent of $|\psi\>$.

{\it Sufficiency.}  Suppose that for each $j\in[n]$, Eq. (\ref{C2.11}) holds with the desired properties. Put
$$\rho_j=\sum_{i=1}^{r}c_{ij}|e_{ij}\>\<e_{ij}|,$$
then $\rho_j\in D(H_j)$ satisfying: $\forall |\psi\>\in Q$,
$$
\tr_{\hat{j}}[S|\psi\>\<\psi|S^\dag]=\tr_{\hat{1}}[T_{j}S|\psi\>\<\psi|S^\dag T_{j}]=\sum_{i=1}^{r}c_{ij}|e_{ij}\>\<e_{ij}|=\rho_j.$$
Thus, $Q$ is masked by  $S$ using Definition 2.1. The proof is completed.

Some relationships between quantum
secret sharing schemes   and quantum error-correcting
codes  were explored in \cite[Theorem7]{[24]}.
The quantum erasure channel (QEC) was considered in \cite{QEC} and pointed out that QECCs for a QEC have to correct for erasures, i.e., arbitrary errors at known positions. It was also proved there that four qubits are necessary and sufficient to encode one qubit and correct one erasure.
Next, let us explore the relationship between QMMs and QECCs.

To do this, we recall some concepts and known results.
According to \cite{QEC}, a quantum channel $\Phi$ on $\mathcal{H}^{(n)}$ is said to a quantum one-erasure channel  if it has Kraus operators $\{E_k\}_{k=1}^m$ of the form
$E_k=T_{j}\left(A_{jk}\otimes \left(\otimes_{i\ne j}I_{H_i}\right)\right)T_{j}^\dag$ for all $k\in[m]$ and for some $j\in[n]$ depending only on $\Phi$, where $A_{jk}$ are operators acting on $H_j$. That is, errors occur only at the $j$th
position  of system. We call such a channel $\Phi$  a {\it $j$-erasure channel}.
 A non-zero subspace $V$ of $\mathcal{H}^{(n)}$ is said to be an {\it quantum error-correcting code (QECC)} \cite{Choi} of a quantum channel ${\mathcal{E}}$ of $\mathcal{H}^{(n)}$ if there exists a quantum channel ${\mathcal{R}}$  of $\mathcal{H}^{(n)}$ such that $$(\mathcal{R}\circ\mathcal{E})({|v\rangle}{\langle v|}))={|v\rangle}{\langle v|},\ \  \forall |v\>\in V.$$

 An operator $E$ on $\mathcal{H}^{(n)}$ is said to be a {\it $j$-erasure operator} if it can be written as $E=T_{j}\left(A_j\otimes\left(\otimes_{i\ne j}I_{H_i}\right)\right)T_{j}^\dag$ for some operator $A_j$ acting on the Hilbert space $H_j$. Thus, a channel $\Phi$ is a $j$-erasure channel if and only if it has Kraus operators consisting of $j$-erasure operators.

 The following conclusion was pointed out in \cite{QEC}.

{\bf Lemma 3.1.} {\it For a given index $j$ in $[n]$,
a non-zero subspace $V$ of $\mathcal{H}^{(n)}$ is a QECC of any $j$-erasure channel ${\mathcal{E}}$ of $\mathcal{H}^{(n)}$  if and only if for every  $j$-erasure operator $E$, the following two conditions are satisfied:}
\be\label{ECCS1}
\<x|E|x\>=\<y|E|y\>,\ \ \forall |x\>,|y\>\in PS(V);
\ee
\be\label{ECCS2}
\<x|E|y\>=0,\ \ \forall |x\>,|y\>\in PS(V),\<x|y\>=0.
\ee

{\bf Lemma 3.2.} {\it Let $|\psi_1\>,|\psi_2\>\in PS(K)$ with $\<\psi_1|\psi_2\>=0$. If the set  \be\label{TT2.2}
Q=\left\{|\psi_1\>,|\psi_2\>,
\frac{1}{\sqrt{2}}(|\psi_1\>+|\psi_2\>),\frac{1}{\sqrt{2}}(|\psi_1\>-\i|\psi_2\>)\right\}
\ee
can  be masked by a linear operator $S: K\rightarrow \H^{(n)}$,
then
\be\label{L2-1}
\tr_{\hat{j}}[S|\psi_1\>\<\psi_2|S^\dag]=0(\forall j\in[n]).
\ee
}

{\bf Proof.} Put
$$|\psi_3\>=\frac{1}{\sqrt{2}}(|\psi_1\>+|\psi_2\>),\ |\psi_4\>=\frac{1}{\sqrt{2}}(|\psi_1\>-\i|\psi_2\>).$$
Since $Q$ is masked by $S: K\rightarrow \H^{(n)}$, we see from Definition 2.1 that there exist states $\rho_j\in D(H_j)$ such that
$$
 \tr_{\hat{j}}[S|\psi_k\>\<\psi_k|S^\dag]=\rho_j, \forall j\in[n], \forall k=1,2,3,4.$$
Let $(t_1,t_2)=\left(\frac{1}{\sqrt{2}},\frac{1}{\sqrt{2}}\right)$ or
$(t_1,t_2)=\left(\frac{1}{\sqrt{2}},\frac{-\i}{\sqrt{2}}\right)$. Then
$|\psi\>:=t_1|\psi_1\>+t_2|\psi_2\>=|\psi_3\>$ or $|\psi_4\>$, and so
\begin{eqnarray*}
\rho_j&=&\tr_{\hat{j}}[S|\psi\>\<\psi|S^\dag]\\
&=&\rho_j+t_1{t_2^*}\tr_{\hat{j}}[S|\psi_1\>\<\psi_2|S^\dag]
+{t^*_1}{t_2}\tr_{\hat{j}}[S|\psi_2\>\<\psi_1|S^\dag].
\end{eqnarray*}
Thus,
$$t_1t_2^*\tr_{\hat{j}}[S|\psi_1\>\<\psi_2|S^\dag]
+{t_1^*}{t_2}\tr_{\hat{j}}[S|\psi_2\>\<\psi_1|S^\dag]=0.$$
This shows that
$$\left\{\begin{array}{l}
         \tr_{\hat{j}}[S|\psi_1\>\<\psi_2|S^\dag]
+\tr_{\hat{j}}[S|\psi_2\>\<\psi_1|S^\dag]=0,\\
         \tr_{\hat{j}}[S|\psi_1\>\<\psi_2|S^\dag]
-\tr_{\hat{j}}[S|\psi_2\>\<\psi_1|S^\dag]=0.
       \end{array}
\right.
$$
Thus, $\tr_{\hat{j}}[S|\psi_1\>\<\psi_2|S^\dag]=0.$ The proof is completed.

Next theorem shows that a linear operator is a universal masker if and only if its range is a quantum error-correcting code of any  one-erasure channel.

{\bf Theorem 3.2.} {\it Let $n\ge2$ and $S:K\rightarrow \mathcal{H}^{(n)}$ be a linear operator. Then $S$ is a universal masker if and only if the range $V={\rm{ran}}(S)$ of $S$ is a quantum error-correcting code of any  one-erasure channel of $\mathcal{H}^{(n)}$.}

{\bf Proof.} {\it Necessity.} Suppose that $S$ is a universal masker with the marginal states $\rho_j\in D(H_j)(j\in[n])$. Let $j\in[n]$. Then Theorem 3.1 yields that
 there exists a PD $\mathcal{P}_j=\{c_{ij}\}_{i=1}^{r}$ with $c_{ij}>0$, and an orthonormal set $\mathcal{E}_j=\{|e_{ij}\>\}_{i=1}^{r}\subset PS(H_j)$ such that $\forall |\psi\>\in PS(K)$,
\be\label{C2.13}
T_{j}S|\psi\>=\sum_{i=1}^{r}\sqrt{c_{ij}}|e_{ij}\>|f^\psi_{ij}\>\in H_j\otimes\left(\otimes_{k\ne{j}}H_k\right),\ee
where $\mathcal{F}^\psi_j=\{|f_{ij}^{\psi}\>\}_{i=1}^{r}$ is an orthonormal set in  $\otimes_{k\in \hat{j}}H_k$ for every $|\psi\>\in PS(K)$.
Extending $\{|e_{ij}\>\}_{i=1}^{r}$ as an ONB $\{|e_{ij}\>\}_{i=1}^{d_j}$ for $H_j$, then every operator $A_j$ on $\H_j$ can be written as
$A_j=\sum_{i,k=1}^{d_j}a^{(j)}_{ik}|e_{ij}\>\<e_{kj}|.$ Thus,  every $j$-erasure operator $E$ can be written as
\be\label{EEE}
E=T_{j}^\dag\left(A_j\otimes\left(\otimes_{i\ne j}I_{H_i}\right)\right)T_{j}=\sum_{i,k=1}^{d_j}a^{(j)}_{ik}E_{i,k},\ee
where
$$E_{i,k}=T_{j}^\dag\left((|e_{ij}\>\<e_{kj}|)\otimes
\left(\otimes_{k\ne j}I_{H_k}\right)\right)T_{j}.$$
For any two  stats  $|x\>$ and $|y\>$ in $K$, we compute from Eq. (\ref{C2.13}) that
\begin{eqnarray*}
\<x|S^\dag E_{i,k}S|y\>&=&
\<x|S^\dag T_{j}^\dag\left((|e_{ij}\>\<e_{kj}|)\otimes\left(\otimes_{i\ne j}I_{H_i}\right)\right)T_{j}S|y\>\\
&=&\sum_{a,b=1}^{r}\sqrt{c_{aj}c_{bj}}
\<e_{aj}|e_{ij}\>\<e_{kj}|e_{bj}\>\<f^x_{aj}|f^{y}_{bj}\>.
\end{eqnarray*}
Hence,
\be\label{313}
\<x|S^\dag E_{i,k}S|y\>=\left\{\begin{array}{cc}
            \sqrt{c_{ij}c_{kj}}\<f^x_{ij}|f^{y}_{kj}\>, & i,k\le r; \\
            0, & \min\{i,k\}>r.
          \end{array}\right.
\ee
Especially,
$$\<x|S^\dag E_{i,k}S|x\>
=\left\{\begin{array}{cc}
            \sqrt{c_{ij}c_{kj}}\delta_{i,k}, & i,k\le r; \\
            0, & \min\{i,k\}>r,
          \end{array}\right.$$
and therefore Eq. (\ref{EEE}) yields that
$$\<x|S^\dag ES|x\>
=\sum_{i,k=1}^{r}a^{(j)}_{ik}\sqrt{c_{ij}c_{kj}}\delta_{i,k},$$
which is independent of the choice of $|x\>$ in $PS(K).$ Hence, Eq. (\ref{ECCS1}) holds.

From Eq. (\ref{EEE}), we obtain that
\begin{eqnarray*}
\tr_{\hat{j}}[S|x\>\<y|S^\dag]&=&\tr_{\hat{1}}[T_{j}S|x\>\<y|S^\dag T_{j}^\dag]\\
&=&\sum_{a,b=1}^{r}\sqrt{c_{aj}c_{bj}}|e_{aj}\> \<e_{bj}|\cdot \<f^y_{bj}|f^x_{aj}\>.
\end{eqnarray*}
Thus,
$$\<e_{ij}|\tr_{\hat{j}}[S|x\>\<y|S^\dag]|e_{kj}\>
=\left\{\begin{array}{cc}
            \sqrt{c_{ij}c_{kj}}\<f^y_{kj}|f^{x}_{ij}\>, & i,k\le r; \\
            0, & \min\{i,k\}>r,
          \end{array}\right.$$
and it follows from (\ref{313}) that
\be\label{E3}
\<x|S^\dag E_{i,k}S|y\>
=\left\{\begin{array}{cc}
            \<e_{ij}|\tr_{\hat{j}}[S|y\>\<x|S^\dag]|e_{kj}\>^*, & i,k\le r; \\
            0, & \min\{i,k\}>r.
          \end{array}\right.
\ee
Thus, when $\<x|y\>=0$, Lemma 3.2 yields that $\tr_{\hat{j}}[S|y\>\<x|S^\dag]=0$ and then Eq. (\ref{E3}) implies
$\<x|S^\dag E_{i,k}S|y\>=0$ and therefore, $\<x|S^\dag ES|y\>=0.$ Thus, Eq. (\ref{ECCS2}) holds.
It follows from Lemma 3.1 that $V={\rm{ran}}(S)$ is a quantum error-correcting code of any  $j$-erasure channel of $\mathcal{H}^{(n)}$.

{\it Sufficiency.} Suppose that the $V={\rm{ran}}(S)$ is a quantum error-correcting code of any  one-erasure channel of $\mathcal{H}^{(n)}$.
 To show that $S$ is a universal masker, it suffices to check that for each $j\in[n], \forall |x\>,|y\>\in PS(K)$ and $\forall i,k\in[d_j]$, it holds that
\be\label{T33M}
\<e_{ij}|\tr_{\hat{j}}[S|x\>\<x|S^\dag]|e_{kj}\>
=\<e_{ij}|\tr_{\hat{j}}[S|y\>\<y|S^\dag]|e_{kj}\>,
\ee
where $\{|\e_{ij}\>\}_{i=1}^{d_j}$ is an ONB for $H_j$.
Let $j\in[n]$. Then $V$ is a quantum error-correcting code of any  $j$-erasure channel of $\mathcal{H}^{(n)}$.
Using Lemma 3.1 yields that
\be\label{T33S}
\<x|S^\dag ES|x\>=\<y|S^\dag ES|y\>, \ \forall |x\>,|y\>\in PS(K)
\ee
for all $j$-erasure operators $E$ on $\H^{(n)}$. Let $|x\>,|y\>\in PS(K)$ and $i,k\in[d_j]$. Put  $$E_{ik}=T_{j}^\dag\left((|e_{ij}\>\<e_{kj}|)\otimes
\left(\otimes_{k\ne j}I_{H_k}\right)\right)T_{j},$$ we obtain that
\begin{eqnarray*}
&&\<e_{ij}|\tr_{\hat{j}}[S|x\>\<x|S^\dag]|e_{kj}\>\\
&=&\tr\left(\tr_{\hat{j}}[S|x\>\<x|S^\dag]\cdot|e_{kj}\>\<e_{ij}|\right)\\
&=&\tr\left(\tr_{\hat{1}}[T_{j}S|x\>\<x|S^\dag T_{j}^\dag]\cdot|e_{kj}\>\<e_{ij}|\right)\\
&=&\tr\left([T_{j}S|x\>\<x|S^\dag T_{j}^\dag]\cdot[|e_{kj}\>\<e_{ij}|\otimes
\left(\otimes_{k\ne j}I_{H_k}\right)]\right)\\
&=&\<x|S^\dag E_{ik}S|x\>.\\
\end{eqnarray*}
Similarly, $$\<e_{ij}|\tr_{\hat{j}}[S|y\>\<y|S^\dag]|e_{kj}\>=\<y|S^\dag E_{ik}S|y\>.$$
Using Eq. (\ref{T33S}) for $E=E_{ik}$ implies Eq. (\ref{T33M}).
 The proof is completed.

It was proved in \cite[Theorem 5]{QEC} that there is no quantum error-correcting codes of length $3$ that can correct one-erasure and encode one qubit. Combining this result with Theorem 3.2, we have the following conclusion, which gives a negative answer to Question 2.1 for the case where $d=2$.

{\bf Theorem 3.3.} {\it There is no  maskers $S:\C^2\rightarrow \C^2\otimes\C^2\otimes\C^2.$ That is, it is impossible to mask all states of $\C^2$ into $\C^2\otimes\C^2\otimes\C^2$.}

To discuss the answer to Question 2.2, we assume that there exists a universal masker $S:K\rightarrow H_1\otimes H_2$. Then the set
$Q$
in Lemma 3.2 is masked by $S: K\rightarrow \H^{(2)}$. Thus,
Theorem 3.1 implies that there exists a  PD $\{c_i\}_{i=1}^{r}$ with $c_i>0$ for all $i\in[r]$ and an orthonormal set $\{|e_i\>\}_{i=1}^{r}\subset PS(K)$ such that
$$S|\psi_k\>=\sum_{i=1}^{r}\sqrt{c_i}|e_i\>|f^{\psi_k}_i\>\ (k=1,2,3,4),
$$
where $\<f^{\psi_k}_s|f^{\psi_k}_t\>=\delta_{s,t}$.
Lemma 3.2 yields that
$$
\tr_{1}[S|\psi_1\>\<\psi_2|S^\dag]=\tr_{2}[S|\psi_1\>\<\psi_2|S^\dag]=0.
$$
Thus,
\begin{eqnarray*}
0&=&\tr_1[S|\psi_1\>\<\psi_2|S^\dag]\\
&=&\tr_1\left[\sum_{i,j=1}^r\sqrt{c_ic_j}|e_i\>\<e_j|\otimes|f^{\psi_1}_i\>\<f^{\psi_2}_j|\right]\\
&=&\sum_{i=1}^rc_i|f^{\psi_1}_i\>\<f^{\psi_2}_i|.
\end{eqnarray*}
Since $\<f^{\psi_2}_i|f^{\psi_2}_1\>=\delta_{i,1}$, we get
$$0=\left(\sum_{i=1}^rc_i|f^{\psi_1}_i\>\<f^{\psi_2}_i|\right)
|f^{\psi_2}_1\>
=c_1|f^{\psi_1}_1\>,$$
a contradiction.

This leads to the following result, which gives a negative answer to Question 2.2.

{\bf Theorem 3.4.}(Generalized no-masking theorem) {\it For given quantum systems described by the Hilbert spaces  $K, H_1,H_2$, there does not exist a universal masker $S: K\rightarrow H_1\otimes H_2$.}

For the answer to Question 2.1 in the case where $d=6$, we have the following, which is a special case of Theorem 3.2.

{\bf Corollary 3.1.} {\it There is a universal masker $S:\C^6\rightarrow\C^6\otimes\C^6\otimes\C^6$ if and only if there is  a quantum error-correcting code $V\subset \C^6\otimes\C^6\otimes\C^6$ of dimension $6$ that can correct the errors of any  one-erasure channel of $\C^6\otimes\C^6\otimes\C^6$.}

{\bf Corollary 3.2.} {\it If there exists a quantum error-correcting code $V\subset\mathcal{H}^{(n)}$ of any  one-erasure channel, then the pure states of any system $K$ of dimension  $\dim(V)$ can be masked   into $\mathcal{H}^{(n)}$.}

\section{Summary and Conclusions}

In this paper, we have first introduced the concept of a quantum multipartite masker (QMM), which is an operator $S$ that maps the pure states of a system $K$ into a multipartite system $\H^{(n)}=\otimes_{k=1}^nH_k$ such that the image states have the same marginal states at each subsystem. Based on the definition, we have derived some basic results and proposed some questions discussed later. Then we have obtained
an expression of a QMM $S$ for a set $Q$ of pure states of a system $K$ and proved that a linear operator $S$ from $K$ into $\H^{(n)}$ is a QMM of all pure states of $K$ if and only if its range $S(K)$ is a quantum error-correcting code (QECC). Thus, all pure states of a system $K$ can be masked   into $n$-partite system $\H^{(n)}$  if and only if $\H^{(n)}$ contains a $\dim(K)$-dimensional QECC. Lastly, we have shown that for given systems $K, H_1,H_2$, there does not exist a universal masker $S: K\rightarrow H_1\otimes H_2$, which generalizes the  no-masking theorem in \cite{Mask}.

As an application, we have proved that there is no universal maskers from $\C^2$ into $\C^2\otimes\C^2\otimes\C^2.$ This conclusion gives a consummation to the Li and Wang's in \cite{Li-Wang} where the case $d=2$ was not considered.  Consequently, the system $\C^3$ can not be masked into $\C^2\otimes\C^2\otimes\C^2$ by a linear operator $S$; otherwise,  $SJ_{2,3}$ would be a universal masker of $\C^2$ into $\C^2\otimes\C^2\otimes\C^2$.
At the end of paper \cite{Li-Wang}, the authors asked:
can all quantum states of level $d$ be hidden into
tripartite quantum system $\C^n\otimes\C^n\otimes\C^n$ with $n<d$ or not? Our conclusion above gives a negative answer to this question for the case where $d=3$.

Furthermore, let $S_d$ be the masker (\ref{Sd}) of $\C^d$ into $\C^d\otimes\C^d\otimes\C^d$ given by \cite{Li-Wang}, where $d\ne2,6$, and let
$J_{d,d+1}$ be the canonical imbedding of $\C^d$ into $\C^{d+1}$, i.e., $J_{d,d+1}|x\>=|x\>\oplus0$. Then we can see from Theorem 2.1 and Corollary 2.1 that for any $d\ge2$, arbitrary states of $\C^d$ can be masked into the tripartite system $\C^{d+1}\otimes\C^{d+1}\otimes\C^{d+1}$ by using the isometric maskers
$\tilde{S}_d=S_{d+1}J_{d,d+1}(d\ne5)$ and $\tilde{S}_5=(J_{5,6}\otimes J_{5,6}\otimes J_{5,6})S_5$, with the same marginal states $\frac{1}{d+1}I_{d+1}(d\ne5)$ and $\frac{1}{5}J_{5,6}^\dag I_5J_{5,6}=\frac{1}{5}I_5\oplus0$, respectively. Thus, we see that
an arbitrary quantum state of $\C^d$ can be encoded into the correlations between any
two subsystems of  of the tripartite system $\C^{d+1}\otimes\C^{d+1}\otimes\C^{d+1}$ by using the masker $\tilde{S}_d$, with none of the information about that state accessible from one subsystem alone. This is a quantum analogue to the one-time pad. Interestingly, such a
quantum analogue is impossible for any pure-state encoding
into two subsystems \cite{[24],Brau-Pati}. This impossibility is also revealed by the generalized no-masking theorem (Theorem 3.4), just like two persons play together with a playing cards is improper and uninteresting.
Another application is that
arbitrary quantum states  of $\C^d$ can be completely hidden
 in correlations between any two subsystems of the tripartite system $\C^{d+1}\otimes\C^{d+1}\otimes\C^{d+1}$, while arbitrary quantum states cannot be completely hidden in the correlations between subsystems of a bipartite system (no-hiding theorem) \cite{Brau-Pati,hiding[7]}.

Moreover, by using Theorem 3.2, we know that for each $d\ge2$,
$V=\tilde{S}_{d+1}(\C^d)$ is a QECC of dimension $d$ for any  one-erasure channel of the system $\C^{d+1}\otimes\C^{d+1}\otimes\C^{d+1}$. For example,
we obtain a QECC $V=\tilde{S}_6(\C^5)$ of dimension $5$ for any  one-erasure channel of the system $\C^{6}\otimes\C^{6}\otimes\C^{6}$, which is generated by the ONB:
$$\tilde{S}_6|k\>=\frac{1}{\sqrt{5}}
\sum_{j=1}^5(|k\>\oplus0)\otimes(|v_{jk}\>\oplus0)\otimes(|w_{jk}\>\oplus0)
(k=1,2,\ldots,5),$$
where $\{|1\>,|2\>,\ldots,|5\>\}$ is an orthonormal basis (ONB) for $\C^5$, $V=[v_{jk}]$ and $W=[w_{jk}]$ are a pair of orthogonal Latin squares of order $5$.

An open question is remained:

{\it Is there a
QECC of dimension $6$ for any  one-erasure channel of the system $\C^{6}\otimes\C^{6}\otimes\C^{6}$?}

\section*{Acknowledgements}
{This subject was supported by the National Natural Science Foundation of China (Nos. 11871318,11771009), the Fundamental Research Funds for the Central Universities (GK202007002, GK201903001) and the Special Plan for Young Top-notch Talent of Shaanxi Province(1503070117)}




\end{document}